\documentclass[%
 12pt,preprint
]{aastex62}
\usepackage{natbib}
\usepackage{graphicx}
\usepackage{dcolumn}
\usepackage{bm}
\usepackage{xcolor}
\usepackage[utf8]{inputenc}
\usepackage[T1]{fontenc}
\usepackage{float}
\usepackage{hyperref}

\newcommand{\bfx}{\mbox{\boldmath $x$}} 
\newcommand{\bfk}{\mbox{\boldmath $k$}} 

\shortauthors{Y. Luo et al.}
\begin{document}

\title{Big Bang Nucleosynthesis with an Inhomogeneous Primordial Magnetic Field Strength}

\author{Yudong Luo$^{\ast}$}
\affiliation{National Astronomical Observatory of Japan,
2-21-1 Osawa, Mitaka, Tokyo 181-8588, Japan}
\affiliation{Department of Astronomy, Graduate School of Science,
University of Tokyo, 7-3-1 Hongo, Bunkyo-ku, Tokyo 113-0033, Japan}
\affiliation{School of Physics and Nuclear Energy Engineering,
and International Research Center for Big-Bang Cosmology and Element Genesis, Beihang University 37, Xueyuan Rd., Haidian-qu, Beijing 100083 China}
\email{$\ast$ ydong.luo@nao.ac.jp}

\author{Toshitaka Kajino}
\affiliation{National Astronomical Observatory of Japan,
2-21-1 Osawa, Mitaka, Tokyo 181-8588, Japan}
\affiliation{Department of Astronomy, Graduate School of Science,
University of Tokyo, 7-3-1 Hongo, Bunkyo-ku, Tokyo 113-0033, Japan}
\affiliation{School of Physics and Nuclear Energy Engineering,
and International Research Center for Big-Bang Cosmology and Element Genesis, Beihang University 37, Xueyuan Rd., Haidian-qu, Beijing 100083 China}

\author{Motohiko Kusakabe}
\affiliation{National Astronomical Observatory of Japan,
2-21-1 Osawa, Mitaka, Tokyo 181-8588, Japan}
\affiliation{School of Physics and Nuclear Energy Engineering,
and International Research Center for Big-Bang Cosmology and Element Genesis, Beihang University 37, Xueyuan Rd., Haidian-qu, Beijing 100083 China}

\author{Grant J. Mathews}
\affiliation{National Astronomical Observatory of Japan,
2-21-1 Osawa, Mitaka, Tokyo 181-8588, Japan}
\affiliation{Center for Astrophysics, Department of Physics,
University of Notre Dame, Notre Dame, IN 46556, U.S.A.}
\date{\today}
\begin{abstract}
We investigate the effect on the Big Bang Nucleosynthesis (BBN)  from the presence of a stochastic primordial magnetic field (PMF) whose strength is spatially inhomogeneous. We assume a uniform total energy density and a gaussian distribution of field strength. In this case, domains of different temperatures exist in the BBN epoch due to variations in the local PMF. We show that in such case, the  effective distribution function of particle velocities averaged over domains of different temperatures deviates from the Maxwell-Boltzmann distribution. This deviation is related to the scale invariant strength of the PMF energy density $\rho_{\rm Bc}$ and the fluctuation parameter $\sigma_{\rm B}$. We perform BBN network calculations taking into account the PMF strength distribution, and deduce the element abundances as functions of the baryon-to-photon ratio $\eta$, $\rho_{\rm Bc}$, and $\sigma_{\rm B}$. We find that the fluctuations of the PMF reduces the $^7$Be production and enhances D production. We analyze the averaged thermonuclear reaction rates compared with those of a single temperature, and find that the averaged charged-particle reaction rates are very different. Finally, we constrain the parameters $\rho_{\rm Bc}$ and $\sigma_{\rm B}$ from observed abundances of $^4$He and D, and find that the $^7$Li abundance is significantly reduced.  
We also find that if the $\eta$ value during BBN was larger than the present-day value due to a dissipation of the PMF or a radiative decay of exotic particles after BBN or if the stellar depletion of $^7$Li occurred, abundances of all light elements can be consistent with observational constraints. 
\end{abstract}
\section{\label{sec:intro}Introduction}
Light element synthesis in the early universe is well described by the standard model of Big Bang Nucleosynthesis (BBN). A comparison of predicted isotopic abundances with observation is essential to constrain cosmological models and the physical processes during the BBN epoch  \citep{2006NuPhA.777..208F,2007ARNPS..57..463S,Cyburt:2016cr,2017IJMPE..2641001M}. The standard BBN (SBBN) model evolves a network of nuclear reactions among primordial elements (mainly D, $^{3}$He, $^{4}$He and $^{7}$Li) in a space-time characterized by general relativity, while the microphysics is characterized by particle interactions described within the standard model of particle physics  \citep{Bertulani:2016ci}.\\

Theoretical calculations of light element abundances in SBBN are now well defined and precise  \citep{ Cyburt:2016cr,2017IJMPE..2641001M}. The only parameter is the baryon-to-photon ratio ($\eta$), which is now well determined from the power spectrum of the cosmic microwave background temperature fluctuations. For the value of $\eta$ derived from Planck or the WMAP-9yr analysis  \citep{2013ApJS..208...20B,2016A&A...594A..13P}, there is excellent agreement between BBN and the observed primordial abundances of D and $^{4}$He  \citep{2003PhLB..567..227C,Cyburt:2016cr}. However the observed abundance of $^{7}$Li in metal-poor halo stars, \citep{1982A&A...115..357S, 2010A&A...522A..26S} implies Li/H=$1.6\times10^{-10}$ which disagrees with the theoretical prediction by about a factor of 3 (Li/H$=5.1\times10^{-10}$)  \citep{Cyburt:2016cr}.\\

A number of suggestions have been proposed to solve this problem. One is that a better understanding of the diffusive transport may be needed to understand the lithium abundances of the metal-poor halo stars on the Spite plateau \citep{Fu:2015uua}. Others have argued for the existence of a stellar mass-dependent mechanism to deplete stellar lithium \citep{Richard:2005et}. Motivated by recent observations \citep{Piau:2006kb} suggest that the interstellar medium (ISM) is in fact quite dynamic, it has also been suggested that the $^{7}$Li depleted ejecta from massive Population III stars may be mixed inefficiently with the proto-Galactic ISM prior to the formation of the MPH stars of the galactic halo.\\

In addition to explanations from astrophysical processes, it has been proposed that the current uncertainties in the cross-sections of relevant nuclear reactions are at the level of 0.2\% for $^{4}$He, 5\% for D and $^{3}$He and 15\% for $^{7}$Li  \citep{Descouvemont:2004ks}. Therefore, a partial solution from the nuclear reaction side might be possible once more accurate measurements are achieved  \citep{Broggini:2012kj}. Experimentally, a recent measurement of the $^7$Be($n$,$p$)$^7$Li reaction suggests that the final state involving the first excited state of $^7$Li$^{\ast}$ can contribute up to 20\% of the total cross section \citep[][]{private,newdata}. Theoretically, detailed nuclear reaction network calculations up to the CNO cycle have been carried out  \citep{Coc:2011jh,Acoc2017}, as well as a Monte Carlo likelihood analysis to make a rigorous approach of the theoretical BBN nuclear reaction networks  \citep{Iliadis:2016ej}. Those results, however, do not give a solutions to the lithium problem. On the other hand, the possibility of new resonance reactions to destroy $^7$Li such as $^{7}$Be$(\alpha,\gamma)^{11}$C and $^{7}$Be($^3$He,2$p\alpha$)$^4$He will be explored in the near future, although it has been found that these resonances must have unrealistically large decay widths \citep{Chakraborty:2011ju,Civitarese:2013jl,Hammache:2013it}.\\

Beside these, a variety of nonstandard BBN models have also been proposed such as an Inhomogeneous BBN \citep{Applegate:1987hp,1987ApJ...320..439A,1988PhRvD..37.1380F,Kajino:1991jf,Orito:1997wn,Lara:2006es,2017IJMPE..2641003N}, dark matter decay \citep{2013PhLB..718..704K}, sterile neutrinos \citep{Esposito:2000iz,2014PhRvD..90h3519I} and super symmetric particles \citep{2008PhLB..669...46A,2011JPhCS.312d2012K,MK2017}. Those possibilities are discussed in \citet{KurkiSuonio:2017wl,2017IJMPE..2641001M,2018AIPC.1947b0014M}.\\

Recently, non-extensive (non-Maxwellian) statistics (Tsallis statistics) have also been proposed as a solution to the lithium problem  \citep{Hou:2017ed}. In this framework, an extra parameter $q$ characterizes the deviation from a Maxwell-Boltzmann (MB) distribution. When $q=1$, the distribution function is the classical Maxwell-Boltzmann distribution \citep{Bertulani:2013cp}. Several physical sources of the parameter $q$ have been discussed  \citep{Lutz:2003cs,2007EL.....7819001B,Rossani:2009cy,Livadiotis:2010eh,2012arXiv1203.4003P}. Among them is the possibility of entropy and/or temperature fluctuations \citep{Wilk:2000jx,Wilk:2002kj} in the equilibrium state. This motivates the interesting speculation that during the BBN epoch, the background photon energy density or temperature may not have a universal homogeneous value as assumed in previous SBBN studies. Here, we explore this possibility in a phenomenological model whereby sub-horizon isocurvature temperature fluctuations arise from fluctuations in a primordial magnetic field (PMF). Previous studies  \citep{Yamazaki:2012jz} introduced a constant scale invariant (SI) PMF strength within a certain co-moving radius during the BBN epoch. However, as the magnetic field evolves, the strength may not always be homogeneous \citep[e.g.][]{2017PhRvD..96l3525M}. That can affect the temperature on large scales. In section 2, we discuss a primordial stochastic magnetic field and an ansatz for its strength distribution. We take into account this inhomogeneity of the PMF strength in a BBN calculation. In section 3, we discuss its effects on primordial-element abundances and analyze the averaged thermonuclear reactions rates in this inhomogeneous PMF model.\\
\section{\label{sec:Mag_field}Stochastic Magnetic Field}
\subsection{\label{sec:HPMF}Homogeneous magnetic energy density}
The origin and evolution of the galactic magnetic field has been a subject of interest for a number of years. From one view point, the galactic magnetic field might be a fossil remnant of {PMFs amplified through the galactic dynamo process}  \citep{1998PhRvL..81.3575S,2004PhRvD..70l3003B}. Several mechanisms have been proposed to explain the origin of a PMF from early cosmological phase transitions \citep{PhysRevLett.95.121301,Ichiki827,Durrer:2013ec,2016RPPh...79g6901S,2016PhRvD..93d3004Y}. {However, these scenario cannot account for a large scale magnetic field. The co-moving correlation length scale for these models is at most given by the horizon during the phase transition, which is much smaller than a typical galaxy size at the present day. One possible solution of this problem is a super-horizon PMF generated during inflation \citep{1988PhRvD..37.2743T,1993PhRvD..48.2499D,2009JCAP...08..025D}. This kind of magnetic field  is ‘‘frozen-in’’ with the dominant fluids. Previous studies have shown that such a PMF can slightly change the weak reaction rate and the electron-positron distribution function while their main effect is the enhancement of the cosmic expansion rate \citep{1996PhLB..379...73G}. In this sub-section, we first consider this case of a super-horizon scale magnetic field.\\}

 
A statistically homogeneous and isotropic magnetic field must have a two-point correlation function for the co-moving wave vector \citep{2011PhR...505....1K}
\begin{equation} \label{eq:3}
\langle B_i(\bfk)B^i(\bfk')\rangle=2(2\pi)^3P_{\rm [PMF]}(k)\delta{(\bfk-\bfk')}.
\end{equation}
 {Here, as in previous studies, we assume that the power spectrum of the PMF energy density is a power law (PL) spectrum.}
\begin{equation} \label{eq:4}
P_{\rm [PMF]}(k)=Ak^{n_B},
\end{equation}
where $n_B$ is the power-law index. {This PL spectrum is the most common assumption for magnetic fields on cosmological scales.}

One can then derive the normalization coefficient \textbf{$A$} from the variance of the magnetic fields in real space. The co-moving PMF strength $B_{\rm \lambda}$ inside a spherical Gaussian radius should be \citep{Mack:2002ew}
\begin{equation} \label{eq:5}
\langle B^i(\bfx)B_i(\bfx)\rangle|_\lambda=B_{\rm \lambda}^2,
\end{equation}
where $\lambda$ is a typical co-moving length scale for the present-day, usually set as 1 Mpc. Then, applying a Fourier transform to \textbf{$k$} space and integrating this together with a window function, the co-moving strength $B_{\rm \lambda}^2$ becomes 
\begin{eqnarray} \label{eq:6}
    \langle B_i(\bfx)B^i(\bfx)\rangle|_{\rm \lambda}=B_{\rm \lambda}^2=\frac{1}{(2\pi)^6}\int d^3k \int d^3k' \nonumber \\
    \times \exp{(-i\bfx\cdot(\bfk-\bfk'))} \langle B_i(\bfk)B^i(\bfk')\rangle|W^2_{\rm \lambda}(k)|.
\end{eqnarray}
Here, the window function $|W(k)|=\exp{(-\lambda^2\bfk^2/2)}$ is required to constrain large values of the wave number. This means that large spatial scales of the PMF are taken into account while the smallest scales are cut off. {A lower cutoff of the PMFs results from decay of the magnetic field on small scales. Magneto-hydrodynamical (MHD) turbulence generates such a cutoff \citep{Durrer:2013ec}. It has been pointed out \citep{Brandenburg:1996dj} that with random initial conditions for the magnetic field, turbulence can have an inverse cascade that transfers the magnetic energy density from small scales to large scales.}  From Eqs. (\ref{eq:3})$-$(\ref{eq:6}), the final result \citep{Yamazaki:2012jz} for the energy density contributed from a PMF is
\begin{eqnarray}\label{eq:7}
     \langle \rho_{\rm B}\rangle=\frac{\langle B^2\rangle}{8\pi}=\frac{1}{8\pi}\int^{k_{[max]}}_{k_{[min]}} \frac{dk}{k}\frac{k^3}{2\pi^2}P_{\rm [PMF]}(k)\nonumber \\
      =\frac{1}{8\pi}\frac{B_{\rm \lambda}^2}{\Gamma(\frac{n_B+5}{2})}[({\rm \lambda} k_{[max]})^{n_B+3}-({\rm \lambda} k_{[min]})^{n_B+3}],
\end{eqnarray}
where $k_{[max]}$ and $k_{[min]}$ are the maximum and minimum wave numbers, respectively. Their values depend on $\lambda/2\pi$. For example, for an averaged magnetic field strength with $\langle\rho_{\rm B}\rangle=0.2\rho_{\rm rad} $ with $\rho_{\rm rad}$ the radiation energy density after the epoch of $e^{\pm}$ annihilation, the magnetic field energy density would be given by
\begin{equation}\label{eq:1}
\langle\rho_{\rm B}\rangle=0.2 \frac{\pi  g_\ast}{30}T^4=0.2\cdot4.506\hspace{3pt}{\rm g}\hspace{3pt}{\rm cm^{-3}}\Big(\frac{ g_\ast}{3.36264}\Big)\Big(\frac{T}{10^9 \rm K}\Big)^4,
\end{equation}
where $ g_\ast =2+(7/8)\cdot6\cdot(4/11)^{4/3} = 3.36264$ is the effective number of statistical degrees of freedom after the epoch of $e^{\pm}$ annihilation. For $T=2.73 \rm K$ at the present day, $\rho_{\rm Bc}=1.57\times10^{-34}{\rm g\ cm^{-3}}=1.412\times10^{-13} {\rm erg\ cm^{-3}}$. With this amount of magnetic energy, the magnetic field would have a present RMS amplitude of $\langle B^2\rangle_0^{1/2}=1.88\mu \rm G$.\\

We note that this magnitude of the PMF strength is much greater than the upper limit of a few nG on a 1 Mpc co-moving scale inferred from the Planck analysis \citep{2016A&A...594A..19P}. However, in the present analysis, we adopt a lower cutoff of the correlation scale below the large scales that are constrained by the CMB power spectrum. {Previous studies \citep{Durrer:2013ec,Brandenburg:1996dj} pointed out that MHD turbulence can lead to such a cutoff.} {Moreover, the fluid-viscosity due to neutrinos and photons can induce damping of magnetic fields \citep{1998PhRvD..57.3264J, 1998PhRvD..58h3502S}. In fact, the MHD modes with wavelengths smaller than the mean free path of neutrinos and(or) photons are in such a diffusion regime.} Such damping process suggests that a PMF on the smaller scales associated with BBN would dissipate by the time of photon last scattering. Hence, a PMF would only affect the CMB power spectrum via the expansion rate. The Planck constraint of $N_{eff}<3.6 $ (95\% C.L.) is consistent with the upper limit of $\rho_{\rm Bc}/\rho_{\rm tot}<0.2$ adopted here.\\

Previous studies  \citep{2008PhRvD..77d3005Y} used Eq. (\ref{eq:1}) to constrain the ratio of the SI energy density contributed from the PMF based upon the CMB power spectrum. The primordial element abundances can also be computed  \citep{2013PhRvD..88j3011Y,2016PhRvD..93d3004Y} by introducing this amount of extra energy density contribution to the total energy density. The present day co-moving length scale $\lambda=$1 Mpc corresponds to a length of $10^{15}$cm during the BBN epoch, which is well beyond the horizon ($10^{10}-10^{12}$ cm) during BBN. Hence, within the horizon volume, the averaged magnetic energy density mainly affects BBN through the expansion rate
\begin{equation}\label{eq:8}
\Big(\frac{\dot{a}}{a}\Big)^2\equiv H^2=\frac{8\pi G}{3}\rho_{\rm tot} \propto \rho_{\rm \gamma}+\rho_{\rm B};
\end{equation}
\begin{equation}\label{eq:9}
\frac{d\rho}{dt}=-3H\Big(\rho+p\Big),
\end{equation}
where $G$ is the gravitational constant and we use natural units, i.e. $c=1$. The quantities $\rho$ and $p$ are the energy density and the pressure respectively. Since $H^{-1}\propto T_{\rm tot}^{-2}$, the epoch of weak decoupling ($H^{-1}=\tau_{\rm wd}$) occurs when
\begin{equation}\label{eq:12}
(8 \pi G)^{-1}T_{\rm tot}^{-2}  \sim G_F^{-2}T_{\rm \gamma}^{-5}.
\end{equation}
The right hand side results from the fact that the weak-reaction cross sections scale as $G^2_F T_{\rm \gamma}^2$ and the background particle number density is proportional to $T_{\rm \gamma}^3$. Then if the magnetic energy density is included, the left hand side of Eq. (\ref{eq:12}) will be smaller ($T_{\rm \gamma}^{-2}>T_{\rm tot}^{-2}$). This leads to a shorter decoupling time or a higher decoupling temperature $T_{\rm wd}$. Since a larger $T_{\rm wd}$ corresponds to larger n/p ratio, the $^4$He abundance will increase consequently. \\
\begin{figure}[h]
\centering
\includegraphics[scale=0.4]{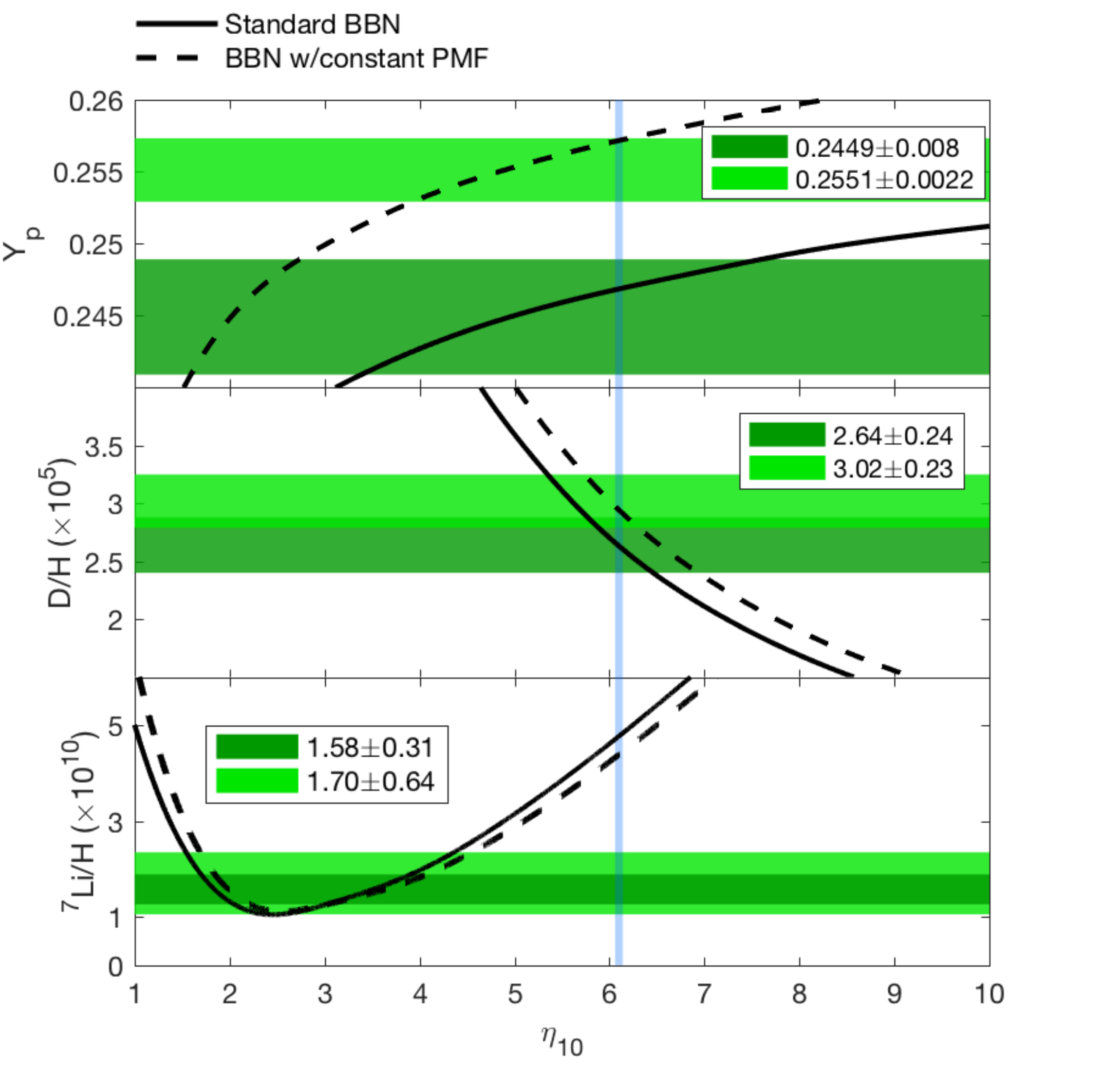}
\caption{\label{fig:epsart} The calculated isotopic abundances in the SBBN model (solid line) and a BBN model with a constant strength of PMF with $\rho_{\rm B}=0.13\rho_{\rm tot}$ (dashed line). The observational values are given by green bands for each isotope. The Planck constraint of $\eta\times10^{10}=6.10\pm0.04$ is given by the vertical blue band. The observed value for each element abundance is given in horizontal painted band. The constraints in the top, middle, and bottom panels are taken from $Y_p$: \citet{2010JCAP...05..003A} (dark-green band),  \citet{Izotov:2014kt} (light-green band); (2)D/H: \citet{Cooke:2014br} (dark-green band),  \citet{2012MNRAS.426.1427O} (light-green band); (3)$^7$Li/H:  \citet{2010A&A...522A..26S}, 1$\sigma$ (dark-green band) and 2$\sigma$ (light-green band) respectively.}
\label{fig1}
\end{figure}
Fig. \ref{fig1} shows the primordial element abundances as a function of $\eta_{10}=\eta\times 10^{10}$. The vertical band shows the limits on the baryon to photon ratio derived from Planck analysis \citep{2016A&A...594A..13P}. The horizontal shaded bands indicate observational constraints on the element abundances.\\

There remain some ambiguities in the primordial abundances \citep[e.g.][]{Cyburt:2016cr}. Hence, in this work, for each element, we list two observational constraints to compare with our calculations: (1)$Y_p$ (mass fraction of $^4$He) : $0.2409-0.2489$ \citep{2010JCAP...05..003A}, or $0.2551-0.2573$ \citep{Izotov:2014kt}; (2)D/H: $2.40-2.88\times 10^{-5}$  \citep{Cooke:2014br}, or $2.79-3.25\times 10^{-5}$ \citep{2012MNRAS.426.1427O}; (3)$^7$Li/H: $1.27-1.89\times 10^{-10}$ \citep[1$\sigma$ from][]{2010A&A...522A..26S}, or $1.06-2.35\times 10^{-10}$ \citep[2$\sigma$ from][]{2010A&A...522A..26S}.\\
 
In Fig. \ref{fig1}, $Y_p$  increases as expected when a PMF is introduced. The other primordial elements D, $^3$He and $^7$Li are also slightly affected. The amount of  PMF energy density is constrained to be $\le13$\% of the total energy density $\rho_{\rm tot}$ in the figure. This is based upon the upper limit to $Y_p$ from the observations of \citet{Izotov:2014kt}. This is equivalent to a co-moving PMF field strength $\langle B^2\rangle^{1/2}=1.51\mu \rm G$.\\
\subsection{\label{sec:IPMF}Inhomogeneous magnetic energy density}
In addition to the effect of a homogeneous PMF energy density, fluctuations of the magnetic field over a wide range of sub-horizon scales will serve as a non-linear driving force that induces the metric fluctuations \citep{Journal:uj}. {As has already been proposed \citep{1997PhLB..390...87D,1999PhRvD..59f3008S,Dolgov:2001kq,2004PhRvD..70l3003B}, it is possible to have an inhomogeneous sub-horizon PMF in the early Universe. Once the turbulence is produced, an induced MHD dynamo can amplify the field exponentially until equipartition between the plasma turbulent kinetic energy and the PMF energy is eventually reached. This can consequently lead to an inhomogeneity in the energy density \citep{Brandenburg:1996dj,2001ApJ...550..824B,2001PhRvE..64e6405C,Dolgov:2001kq}.}\\

For a magnetic field on small scales, the strength can be damped due to photon and neutrino viscosities. This means that the magnetic field on scales with $L<\sqrt{t_{\rm age}(T)\zeta}$ \citep{Durrer:2000jy} dissipates rapidly, where $t_{\rm age}$ is the age of the universe and $\zeta$ is the magnetic diffusivity. An estimate of the damping scale due to the viscosity in the magneto-hydrodynamic evolution process is given by splitting long  and short wavelength fluctuations in the \textbf{B} field separately \citep{Brandenburg:1996dj}. Moreover, a magnetic field with scale $L\ll\sqrt{t_{\rm age}(T)\zeta}$ is not easy to generate, while that with $L\gg\sqrt{t_{\rm age}(T)\zeta}$ will not dissipate, and the magnetic field is frozen-in with the dominant fluids \citep{dendy1990plasma}. Thus, the survival length scale for the PMF during the BBN epoch with temperature set as $0.3$ MeV is $L_{\rm sur} \sim 10^4$ cm  \citep{2012PhR...517..141Y}. This is much smaller than the co-moving length scale for which a constraint on the field amplitude is given from the CMB power spectrum, i.e., $1/(1+z)$ Mpc $\sim 10^{15}$ cm for the BBN redshift of $z\sim 10^9$. Therefore, we cannot exclude the possibility of fluctuations in the PMF length scales of the same order as $L_{\rm sur}$.  We can also consider that the energy density of the PMF could have some distribution $f(\rho_{\rm B})$ rather than the ideal case with $f(\rho_{\rm B})=\delta(\rho_{\rm B}-\rho_{\rm Bc})$. The effect of a PMF on baryons and the $e^+-e^-$ plasma has also been studied \citep{1996PhLB..379...73G,Kawasaki:2012kn}. However, the effect they discussed is not very important for the present application since the modification to the distribution functions is proportional to $ZeB/T^2$. However, if a distribution function $f(\rho_{\rm B})$ exists, then the associated radiation energy density fluctuations can modify the nuclear reaction yields after averaging over all local regions.\\

Most BBN network calculations have considered the photon energy density to be homogeneous during the entire epoch. Here, however, we consider large-scale energy density fluctuations in the temperature (or equivalently photon energy density). The nuclear reactions occur locally, this means that the local velocity distribution function for baryons is, 
\begin{equation}\label{eq:13}
f_{\rm MB}(v|\beta') =\Big(\frac{m \beta'}{2\pi}\Big)^{3/2}4\pi v^2 \exp{(-\frac{\beta' mv^2}{2})}.
\end{equation}
Here, $\beta'$ refers to the inverse temperature $1/kT'$ and $T'$ corresponds to the local temperature. This is just the classical MB distribution which refers to the velocity distribution function of particles for a certain temperature in equilibrium. Since the nucleon gas in the early Universe was dilute, two-body nuclear reactions dominate. The local two-body reaction rate per unit volume can be written as
\begin{equation}\label{eq:22}
R_{12}(\beta')=\frac{N_1N_2}{1+\delta_{12}}\langle\sigma  v\rangle(\beta'),
\end{equation}
where $N_1$ and $N_2$ are the number densities of reacting particles 1 and 2, respectively, $\delta_{12}$ is the Kronecker's delta function for avoiding the double counting of identical particles 1 and 2, and $\langle\sigma  v\rangle(\beta')$ is the averaged thermonuclear reaction rate for a given temperature written as
\begin{equation}\label{eq:23}
\langle\sigma  v\rangle(\beta')=\int \sigma(E)vf_{\rm MB}(v|\beta')dv=\int \Big(\frac{m \beta'}{2\pi}\Big)^{3/2}4\pi v^3\sigma(E)\exp{(-\frac{\beta' mv^2}{2})}dv,
\end{equation}
where $m_{12}$ is the reduced mass of the system 1+2. Because local fluctuations of the energy density occur due to the inhomogeneous PMF, locally nuclei obey a classical MB distribution with inverse temperature equal to $\beta'$. The thermonuclear reaction rates  averaged over the set of temperature fluctuations is then given by
\begin{equation}\label{eq:24}
\langle\sigma  v\rangle(\beta)=\int \langle\sigma  v\rangle(\beta')f(\beta')d\beta'
=\int \Big[\int \sigma(E)vf_{\rm MB}(v|\beta')dv\Big] f(\beta')d\beta'\nonumber 
=\int  \sigma(E)vF(v) dv.
\end{equation}
In the last equation, we defined a new function $F(v)$ which is independent of $\beta'$ as an effective distribution function  averaged over the set of temperature fluctuations\footnote{Note: $F(v)$ is the average velocity distribution function over a length scale much longer than the typical size of magnetic domains' but not a real particle velocity distribution.} In principle, the evolution of nuclear abundances should be solved inhomogeneiously, i.e. the abundance at a given time depends on locations, i.e., Y$_i(t, x)$. But in the present calculation, the inhomogeneity of nuclear abundances is neglected, i.e., Y$_i(t)$. Then, an average distribution function can be defined.
\begin{equation}\label{eq:21}
F(v)\equiv\int d\beta' f(\beta')f_{\rm MB}(v|\beta').
\end{equation}
Here, $f(\beta')$ is the distribution function of $\beta'$ generated from averaging over fluctuations of the energy density. The derivation of this deviation from a classical MB distribution is similar to that deduced in \citet{Beck:2001bq} in terms of Tsallis statistics. Now, we can invoke the central limit theorem and simply assume that the distribution function of magnetic energy density $f(\rho_{\rm B})$ follows a gaussian distribution with a peak located at the mean value $\rho_{\rm Bc}$ ($\langle\rho_{\rm B}\rangle$ in Eq. (\ref{eq:7}))
\begin{equation} \label{eq:16}
f(\rho_{\rm B}) = \frac{1}{\sqrt{2\pi}\sigma^\dagger_{\rm B}}\exp{\Big[-\frac{(\rho_{\rm B}-\rho_{\rm Bc})^2}{2{\sigma^\dagger}_{\rm B}^2}\Big]}.
\end{equation}
We then introduce the fluctuation parameter $\sigma_{\rm B}$ as a dimensionless quantity, i.e., $\sigma_{\rm B}=\sigma^\dagger_{B}/\rho_{\rm Bc}$ to describe the fluctuations of the PMF. In the limit of $\sigma_{\rm B} \to 0$, this is a delta function which corresponds to the homogeneous case. Now we assume that the total energy density is uniform for all volumes, but with some fraction contributed from the magnetic energy density:
\begin{equation}\label{eq:17}
\rho_{\rm tot}=\rho_{\rm B}+\rho_{\rm rad}=const,
\end{equation}
an effective temperature $T_{\rm eff}$ can be defined as
 \begin{equation}\label{eq:18}
\rho_{\rm tot}=\frac{\pi  g_\ast}{30}T_{eff}^4.
\end{equation}
Since $\rho_{\rm tot}$ is constant, the magnetic energy density can not exceed $\rho_{\rm tot}$ in which case $\rho_{\rm rad}$ would obtain an unphysical negative value. Here, we impose a cut-off to the distribution function $f(\rho_{\rm B})$ ($\rho_{\rm B}<0.25\rho_{\rm tot}$).\\
\begin{figure}[h]
\centering
\includegraphics[scale=0.6]{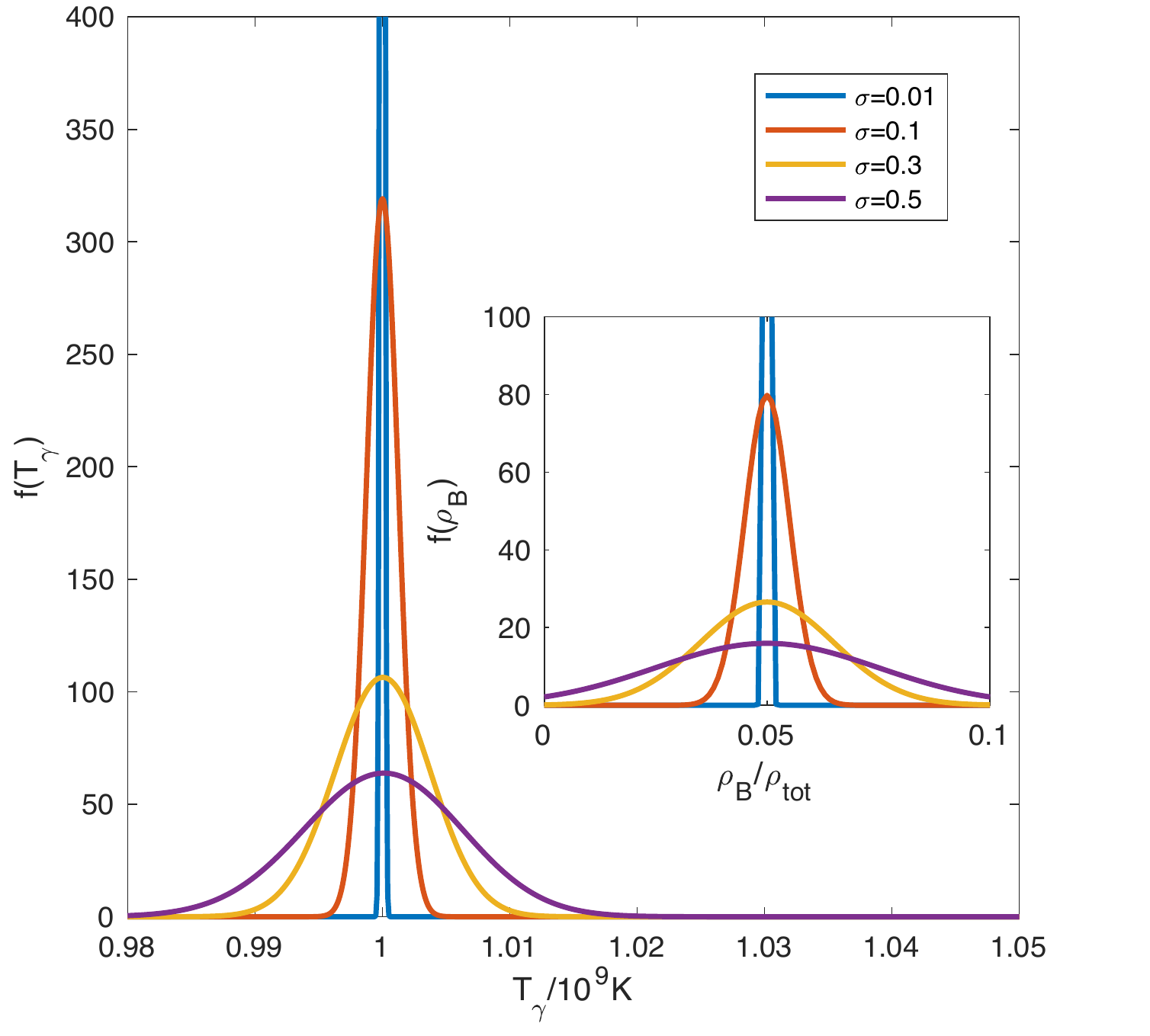}
\caption{\label{fig:epsart} Temperature distribution under the assumption of an inhomogeneous PMF strength. Here $T_{\rm \gamma}$ is in units of $10^9 K$ (centered at $T_9=1$) and $\langle\rho_{\rm B}\rangle$ is taken as 0.05 of $\rho_{\rm tot}$. When $\sigma_{\rm B}<0.01$. the distribution function $f(T_{\rm \gamma})$ can approximately be treated as $\delta(T_{\rm \gamma}-10^9K)$.}
\label{fig2}
\end{figure}
Fig. \ref{fig2} shows Gaussian functions for various values of $\sigma_{\rm B}$. Since we do not expect a very large inhomogeneity in the magnetic energy density strength during BBN, a narrow distribution $f(\rho_{\rm B})$ is required. For $\sigma_{\rm B}<0.65$, $f(\rho_{\rm B})$ is consistent with our cut-off range for $\rho_{\rm B}$.
The photon temperature $T_{\rm \gamma}$ determines the radiation energy density as $\rho_{\rm rad}\propto T_{\rm \gamma}^4$, so Eq. (\ref{eq:17}) becomes
\begin{equation}\label{eq:19}
\beta=1/T_{\rm \gamma}=\Big[T_{\rm eff}^4-\frac{30}{\pi g_\ast}\rho_{\rm B}\Big]^{-1/4}.
\end{equation}
The final expression for the distribution function for $\beta$ is then
\begin{equation}\label{eq:20}
f(\beta)=\frac{1}{\sqrt{2\pi}\sigma_{\rm B}}\exp{\Big[-\frac{(\frac{\pi  g_\ast}{30}(T_{\rm eff}^4-\beta^{-4})-\rho_{\rm Bc})^2}{2\sigma_{\rm B}^2}\Big]}\frac{2\pi  g_\ast}{15}\beta^{-5}.
\end{equation}
\\
\subsection{\label{sec:EffectGW}Effect on reaction rates}
\begin{figure}[h]
\centering
\includegraphics[scale=0.4]{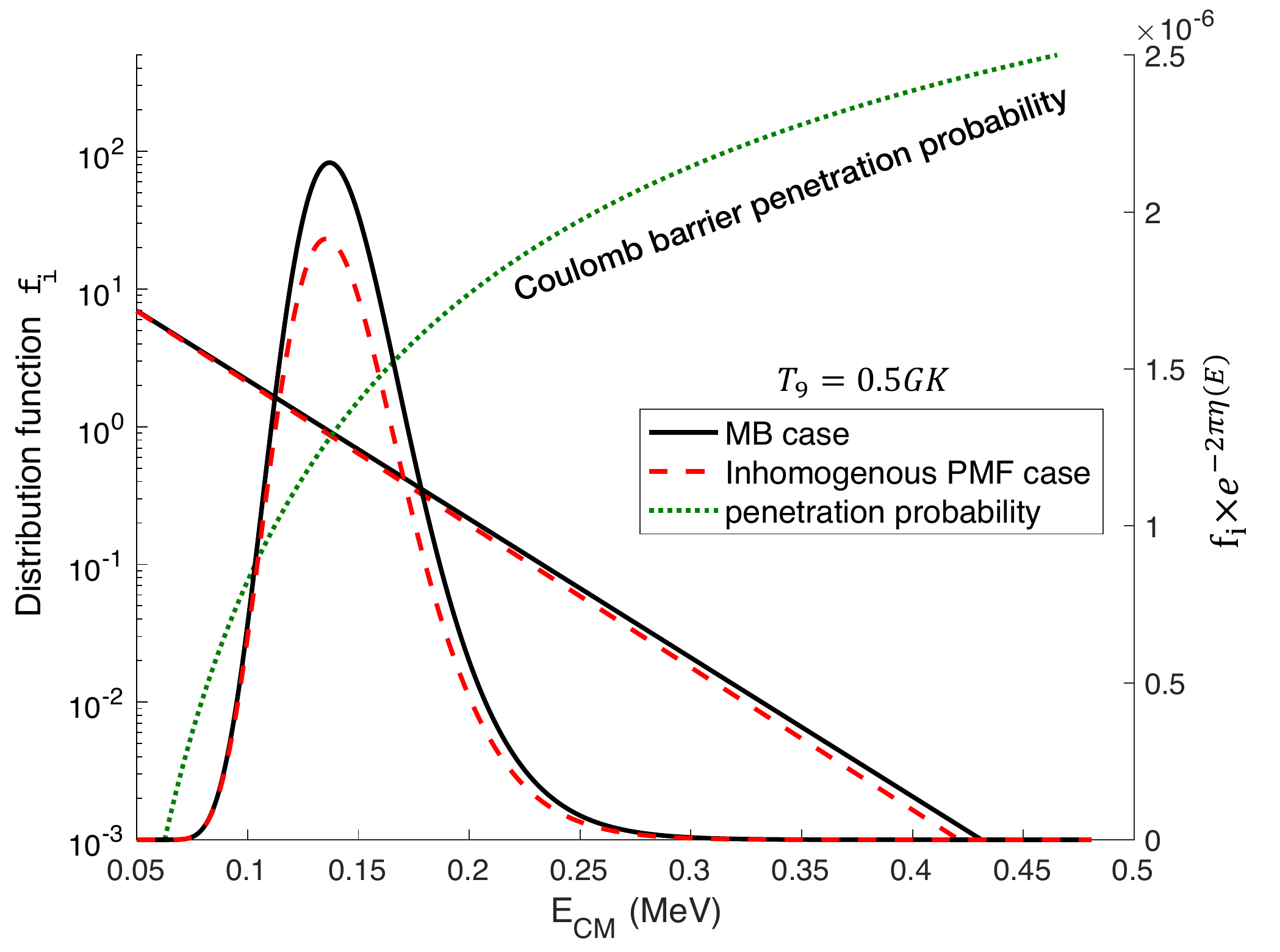}
\caption{\label{fig:epsart} The deviation of the Gamow window for the $^3$He($\alpha$, $\gamma$)$^7$Be reaction in our PMF model from that of the MB case at $t\sim670s$ which corresponds to $T_9 = 0.5$ in SBBN. Although the deviation of the distribution function itself is not large (solid straight line versus the dashed straight line), the Gamow peak in the PMF model (dashed curve) is suppressed compared with the classical Gamow peak for the homogeneous BBN (solid curve).}
\label{fig6}
\end{figure}

Adopting this as the distribution function, we show that the averaged charged particle reactions are affected significantly by the inhomogeneous temperature distribution. For neutron induced reactions, the transmission probability of a neutron through the nuclear potential surface is proportional to the inverse of the velocity $v$ within the assumption of a sharp potential surface \citep{Bertulani:2016ci}. Hence, the cross section is usually expressed as $\sigma(E)_{\rm neutral}=R(E)/v$, where $R(E)$ is a smooth function. Therefore, the change of reaction rates is mainly determined by the deviation of the average distribution function from a MB distribution function. This is not a large effect as shown in Fig. \ref{fig6} (solid straight line and dashed straight line).\\

For charged particle reactions, the astrophysical S-factor is introduced to rewrite the cross section $\sigma(E)$ in terms of a much smoother dependence on the center of mass energy $E$ :
\begin{equation}\label{eq:31}
\sigma(E)_{\rm charged}=\frac{\exp{[-2\pi\eta(E)]}}{E}S(E),
\end{equation}
where $\exp{[-2\pi\eta(E)]}$ approximately expresses the probability to penetrate the Coulomb barrier. This is also known as Gamow factor, $2\pi\eta(E)=\sqrt{E_G/E}$. Eq (\ref{eq:23}) is peaked at the so called Gamow energy $E_G=2m_{12}(\pi eZ_1Z_2)^2$.The deviation from a MB distribution function in the inhomogeneous PMF model is not large. However, the impact on reaction rates can increase when we take into account the factor of $\exp{[-2\pi\eta(E)]}$ for charged particle reactions as shown in Fig. \ref{fig6}. The distribution function (shown by straight lines) in our PMF model looks similar to MB distribution function. However, $\exp{[-2\pi\eta(E)]}$ is also a energy dependent function, and the inhomogeneous PMF model suggests an effective reduction of the Gamow window derived by multiplying this term with the average distribution function $F(v)$.
\\
In conclusion, for the case of an inhomogeneous PMF during BBN epoch, the effect generated from the distribution of PMF energy density can be divided into two parts: 1) changes in the Hubble expansion rate (see Section \ref{sec:HPMF}) and 2) changes within nuclear reaction rates due to an effective non-MB averaged distribution function when we calculate the sum of averaged thermonuclear reaction rates in all domains.\\
\section{\label{sec:Results}Results and Discussion}
\subsection{\label{sec:standard}Standard case}
We have encoded the temperature averaged reaction rates as described in Eqs. (\ref{eq:24}) and (\ref{eq:20}) to calculate the BBN reaction network and compare the results with the observationally inferred abundances for D, $^4$He and $^7$Li. We use the current Particle Data Group world average value $\tau_n=880.3$ s for the neutron lifetime \citep{2014ChPhC..38i0001O}. The baryonic density of the Universe or $\eta$ is now deduced to be $\eta_{10}=6.10\pm0.04$ \citep{2016A&A...594A..13P} from the observations of the anisotropies of the CMB radiation.\\
\begin{figure}[h]
\centering
\plotone{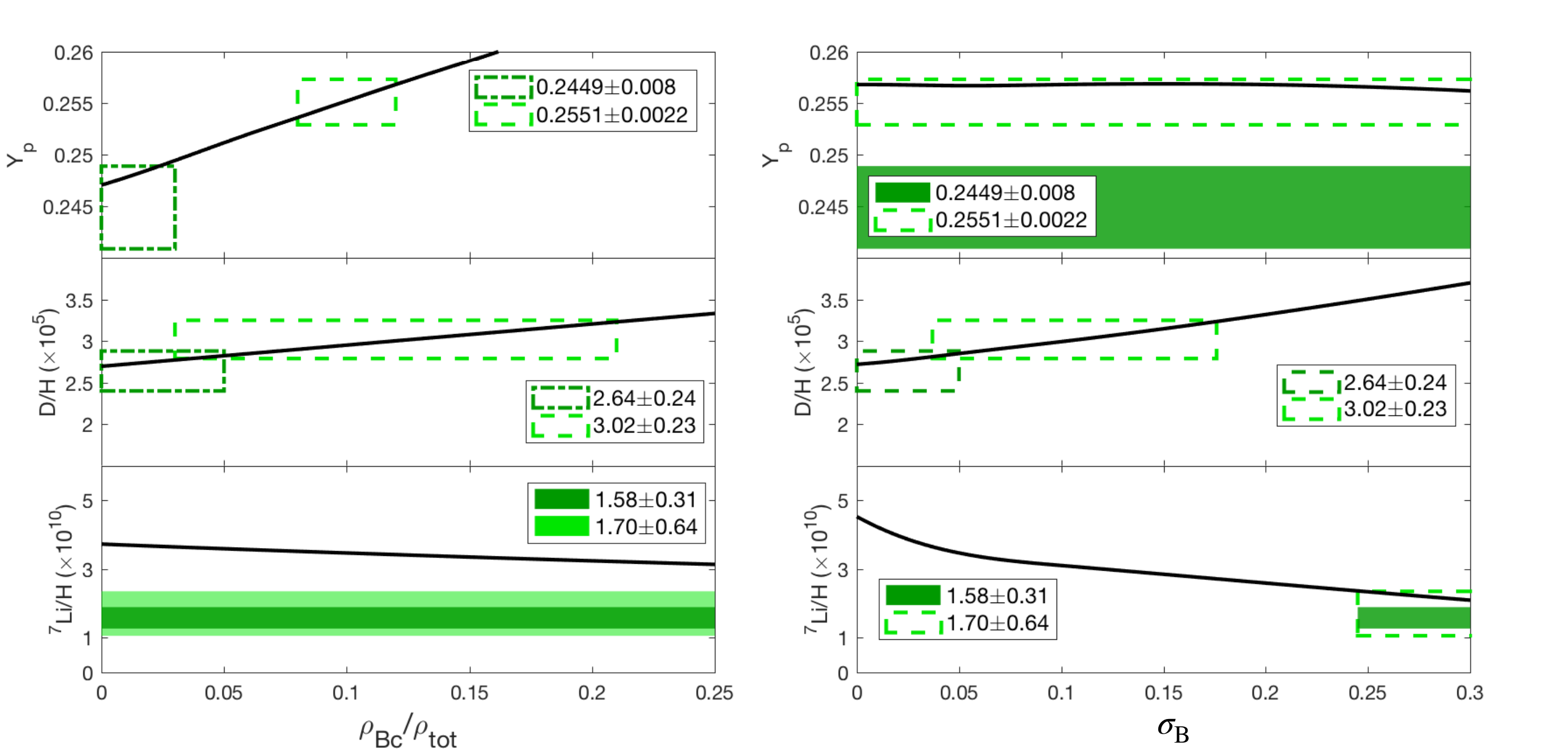}
\caption{\label{fig:epsart}Abundances of $Y_p$ (mass fraction of $^4$He), D/H and $^7$Li/H as a function of $\rho_{Bc}$(left panel) {and $\sigma_{\rm B}$ (right panel)}. {In the left panel,} the fluctuation parameter $\sigma_{\rm B}$ is fixed as 0.05. In right panel, the mean value for the PMF strength is chosen as $1.51\mu G$ thus $\rho_{\rm Bc}/\rho_{\rm tot}=0.13$.  The baryon to photon ratio $\eta$ is set to the best fit value of $\eta_{10}=6.10$ from Planck. {In this figure, both boxes and painted patches refer to the observational constraints on elemental abundances. If the calculated curves have an overlap with observational data, boxes are used. Otherwise painted patches are used.} }
\label{fig3}
\end{figure} 
Fig. \ref{fig3} shows the parameter dependence of the final primordial light element abundances as a function of $\rho_{\rm Bc}$ (left panel) and $\sigma_{\rm B}$(right panel). In the left panel, element abundances are presented as a function of mean magnetic energy density ($\rho_{\rm Bc}$) for a fixed value of $\sigma_{\rm B} = 0.05$. The effect of $\rho_{\rm Bc}$ on the primordial element abundances is consistent with  a PMF model with a homogeneous energy density in the previous study of \citet{Yamazaki:2012jz}: $^4$He is most sensitive to the changes of the cosmic expansion rate, which is equivalent to a change of $\rho_{\rm Bc}$. The constraint from the observed value of $Y_p=0.2551\pm 0.0022$ and D/H implies that the PMF mean energy density has an upper limit of $\rho_{\rm Bc}<0.13\rho_{\rm tot}$.  The right panel shows the element abundances as a function of the fluctuation parameter $\sigma_{\rm B}$. For this panel we set $\rho_{\rm Bc}/\rho_{\rm tot}=0.13$ which is the upper limit from the $Y_p$ observations. In the case that the fluctuation parameter approaches $\sigma_{\rm B} \to 0$ (i.e no fluctuation occurs), the result is consistent with the homogeneous energy density PMF model of \citet{Yamazaki:2012jz}. As $\sigma_{\rm B}$ increases, the inhomogeneity enhances. This affects the element abundances. It is also generally true that as  the D abundance increases, the $^7$Li production is reduced. In this case, the other primordial abundances are strongly dependent on $\sigma_{\rm B}$, while $Y_p$ remains nearly the same as in the case of the homogeneous PMF model. This is a completely new effect on BBN from a PMF model which includes spatial inhomogeneities in the energy density. Finally, from the $Y_p$ and D constraints, we obtain $\rho_{\rm Bc}/\rho_{\rm tot}=0.08-0.13$ and $\sigma_{\rm B}=0.04-0.17$ without violating the observational constraints on the $^4$He and D abundances.\\

\begin{figure}[h]
\centering
\includegraphics[scale=0.4]{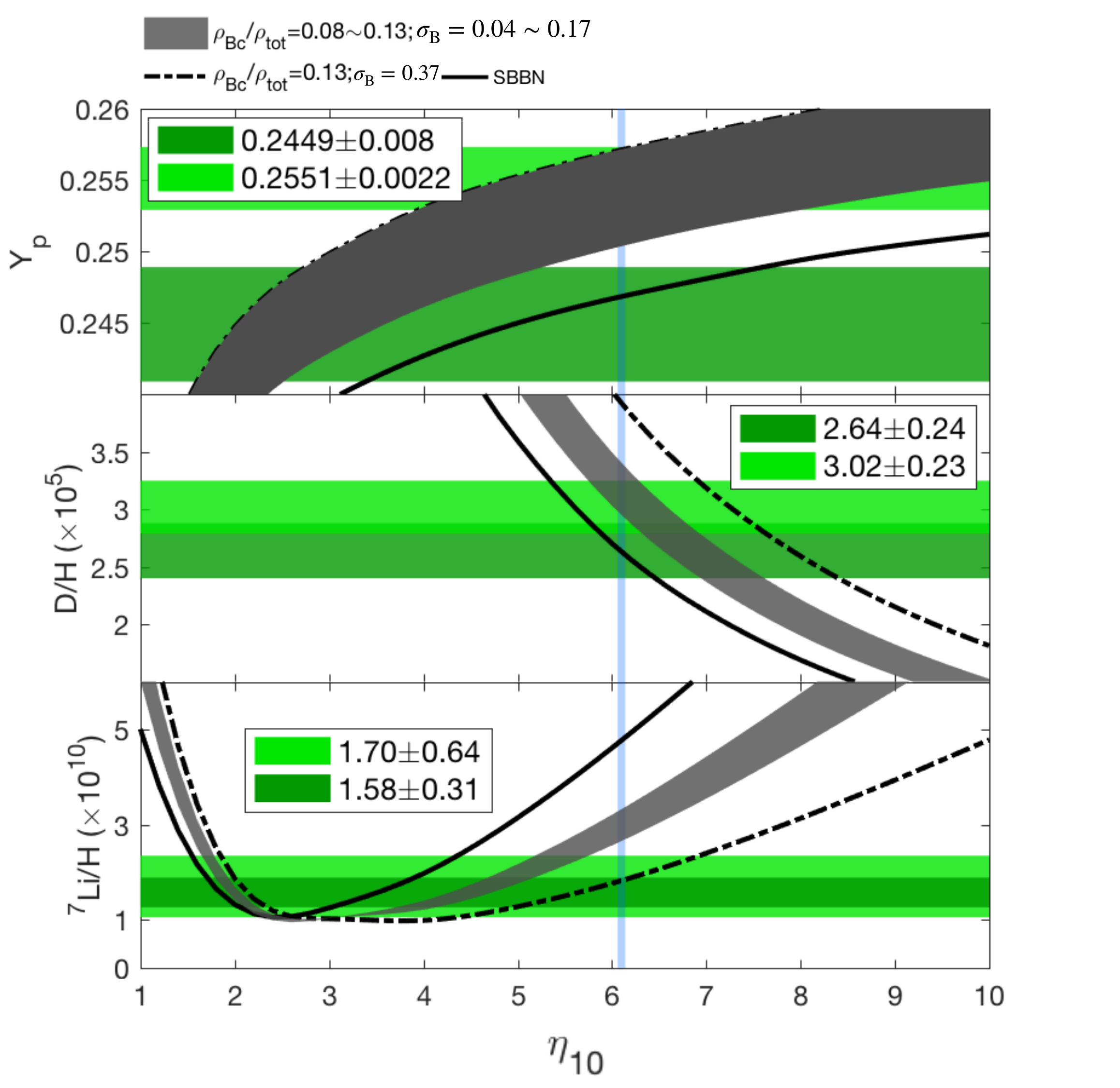}
\caption{\label{fig:epsart} Abundances of $Y_p$, D/H and $^7$Li/H as a function of the baryon to photon ratio $\eta$. The boxes show the adopted observational constraints similar to those in Fig. \ref{fig1}.  This figure shows that larger $\sigma_{\rm B}$ values can suppress the production of $^7$Li but increase the value of D/H. The vertical blue band shows the Planck constraint on $\eta_{10}$.}
\label{fig4}
\end{figure}

In Fig. \ref{fig4}, we illustrate the light element abundances as a function of $\eta_{10}$ with the allowed parameter values of $\rho_{\rm Bc}$ and $\sigma_{\rm B}$. In the grey region, the D/H and $Y_p$ calculations are consistent with observations, and the $^7$Li/H value is reduced to $(3.18- 3.52) \times 10^{-10}$ compared with SBBN. However, this is still above the Spite plateau \citep{1982A&A...115..357S,2010A&A...522A..26S}. The calculated primordial element abundances for $\eta_{10}=6.10$ are shown in Table.\ref{table.1}. Finally, by keeping $\rho_{\rm Bc}/\rho_{\rm tot}=0.13$ which is the upper limit for the mean magnetic energy density, we find that the predicted $^7$Li/H abundance reduces to $1.89\times10^{-10}$ with a fluctuation parameter $\sigma_{\rm B}=0.37$ (dash-dotted line in Fig.\ref{fig4}). Since this parameter region is inside the allowed region of observed $\eta$, the 'Lithium Problem' may be solved in this model. However, the D abundance is D/H$=3.76\times10^{-5}$ which is inconsistent with the observational upper limits \citep{2012MNRAS.426.1427O,Cooke:2014br}.\\

\begin{deluxetable*}{ccCrlc}[h!]
\tablecaption{\label{table.1}Predicted abundances for the BBN primordial light elements ($\eta_{10}=6.10$). Observational data are listed for comparison. For the PMF case, we set $\rho_{\rm Bc}/\rho_{\rm tot}=0.08- 0.13$ based upon the $Y_p$ and D constraints.}
\tablecolumns{8}
\tablenum{1}
\tablewidth{0pt}
\tablehead{
\colhead{Abundance} &
\colhead{SBBN} &
\colhead{PMF with $\sigma_{\rm B}=0.04-0.17$} &
\colhead{Observation}
}
\startdata
\bf{$Y_p$}                       & 0.2469 & $0.2503-0.2536$   & 0.2551$\pm$0.0022 \\
\bf{D/H($\times$10$^{5}$)}       & 2.57     & $2.75-2.96$  &3.02$\pm$0.23 \\
\bf{$^7$Li/H($\times$10$^{10}$)} & 4.91     & $3.18-3.52$  & 1.70$\pm$0.64  \\
\enddata
\end{deluxetable*}

\begin{figure}[h]
\centering
\includegraphics[scale=0.4]{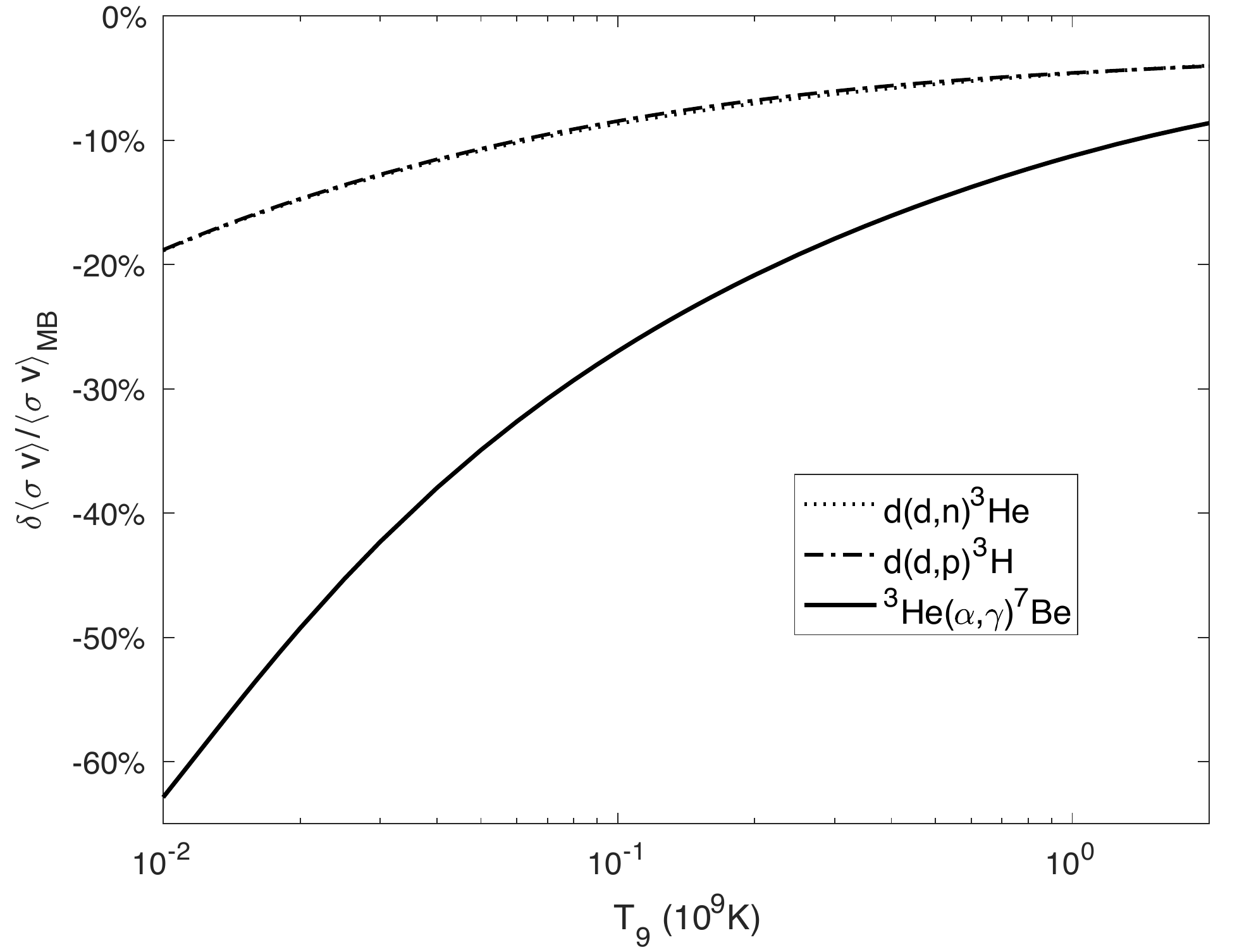}
\caption{\label{fig:epsart} Ratio of reduction in reaction rates of charged particles from an inhomogeneous-strength PMF model compared with the MB case. Here we chose $\sigma_{\rm B}=0.15$.}
\label{fig7}
\end{figure}

The thermonuclear reaction rates are key factors in determining the final primordial abundances. As shown in Fig. \ref{fig4}, D/H  is enhanced and $^7$Li/H reduced as a result of an inhomogeneous PMF energy density model. The $n(p,\gamma)^2$H reaction is the main production mechanism for deuterium, while the $^2$H($d$,$n$)$^3$He and $^2$H($d$,$p$)$^3$H reactions are the main destruction channels. For $^7$Li (or $^7$Be), the main production reaction is $^3$He$(\alpha,\gamma)^7$Be. The main destruction process is $^7$Be$(n,p)^7$Li.  In Fig. \ref{fig7}, we show the reduction fraction for charged particle reaction rates in our PMF model compared with the SBBN results as a function of temperature. For lower temperature, the reduction is larger than that at higher temperature. Since the reaction $^3$He($\alpha,\gamma$)$^7$Be has the largest Coulomb barrier, the reduction is large compared to the D destruction reactions. For deuteron destruction reactions, i.e., $^2$H($d$,n)$^3$He (dotted line)  and $^2$H($d$,$p$)$^3$H (dash-dotted line), we see the same trend in the low energy region since they have the same Gamow energy $E_G$. The solid line shows a larger reduction for the beryllium production rate $^3$He$(\alpha,\gamma)^7$Be at low temperature. Because of the stronger Coulomb repulsion for this reaction, a large $E_G$ contributes to a steeper exponential term for charged particle reaction rates (cf. Fig. \ref{fig3}). Hence, we conclude that all 3 reaction rates which determine the D and $^7$Li abundance are reduced by an inhomogeneous magnetic energy density. \\

\subsection{\label{sec:Nonstandard}Dissipation of PMF and other Effects}
\begin{figure}[H]
\centering
\includegraphics[scale=0.5]{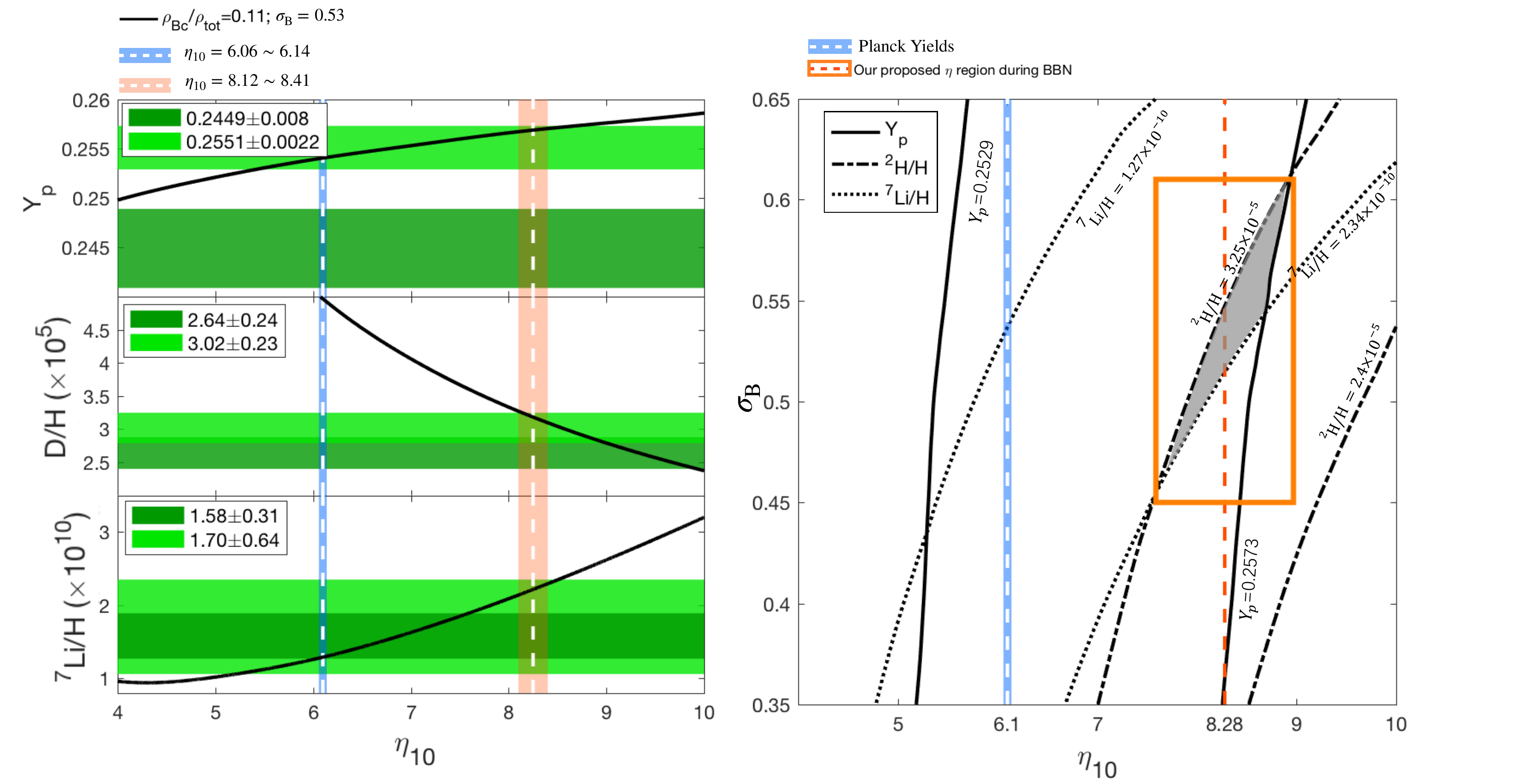}
\caption{\label{fig:epsart} Primordial element abundances as a function of $\eta_{10}$ for fixed $\sigma_{\rm B}=0.53$ (left panel). The horizontal bands show the observational constraints on abundances. The light blue vertical band is the value inferred from the Planck analysis, and the light orange band shows the possible $\eta_{10}$ region for which concordance is possible for all three elements. In the right panel, the contour plot for all three elements is presented. The light blue vertical band is the value from the Planck analysis. In the grey region abundances of all three elements are consistent with observational constraints, and the orange rectangle indicates the constraints on $\eta_{10}$ and $\sigma_{\rm B}$ that are consistent with observational constraints on abundances.}
\label{fig5}
\end{figure}

The above discussion is based on the presumption that no other physical process occurs between BBN and the photon last scattering epoch so that the $\eta$ value from the Planck analysis is the same as that during BBN. In Fig. \ref{fig5} we explore the possibility to find a parameter region with a concordance for all light element abundances with a higher value for the baryon-to-photon ratio. The fraction $\rho_{\rm Bc}/\rho_{\rm tot}$ is chosen as 0.11 which is the mean magnetic field strength constrained from the observed mean $^4$He abundance. In the left panel, the calculated element abundances are shown as functions of $\eta_{10}$ for the fluctuation parameter $\sigma_{\rm B}=0.53$. Although there is no solution to the Li problem within the $\eta_{10}$ range of Planck (light blue vertical band), at $\eta_{10}=8.2\pm0.1$ (light orange vertical band), all of the elements fall into a region that is consistent with the observational constraints. In the right panel we expand this result to a parametric study of the fluctuation parameter $\sigma_{\rm B}$. This panel shows contours for light nuclear abundances in the plane of $\eta_{10}$ and $\sigma_{\rm B}$. The upper limit of $\sigma_{\rm B} \leq 0.65$ satisfies our upper limit on the contribution of the PMF for the case of $\langle\rho_{\rm B}\rangle =0.11 \rho_{\rm tot}$. Here, for a larger fluctuation parameter $\sigma_{\rm B}$ which is taken to be $0.45-0.61$, there is an area (grey-shaded area) in which the abundances of all light elements D, $Y_p$ and $^7$Li are consistent with observations. However, the baryon-to-photon ratio in this region is $\eta_{10}=7.59 - 8.97$, which is larger than the Planck observational constraints (light blue vertical band).\\
We note, that a dissipation of the PMF between BBN and the last scattering of the background radiation could result in an evolution of the $\eta$ value \citep[see Sec. IV-B in][]{Yamazaki:2012jz}. For example, an 11\% increase in the total energy density by the PMF leads to a 13.5 \% increase in the photon number density from the case without dissipative heating. As a result, the $\eta$ value during BBN would be 13.5\% larger than the value after the dissipation. Since this change is not enough to explain the 30\% increase required for the high $\eta$ value in Fig. \ref{fig5} (left panel), the inhomogeneous PMF model alone still cannot completely solve the Li problem.\\

 However, there are other possible astrophysical and cosmological effects that might solve the Li problem: The first is our inhomogeneous PMF model with an amplitude smaller than the best range found in Fig. \ref{fig5} (left panel), coupled with a possible stellar Li depletion. The depletion of $^7$Li during both pre-main sequence \citep{Fu:2015uua} and main sequence phases \citep{Richard:2005et,2006Natur.442..657K} of POP II metal poor stars indicates that the current constraints on the $^7$Li abundance from those stars might be lower than the actual value of primordial $^7$Li abundance. In such a case, the PMF effects on BBN and its dissipation could be a solution to the Li Problem. \\
 
The second possible effect is a change in the $\eta$ value induced by the radiative decay of exotic particles \citep{1982PhRvD..25.1481K,1988ApJ...331...19S,1988ApJ...331...33S,2003PhRvD..68f3504F,2014PhRvD..90h3519I}, which is independent of the dissipation of the PMF as discussed above. In this case, the 30\% increase of the baryon-to-photon ratio in Fig. \ref{fig5} (left panel) might be acceptable.
\begin{deluxetable*}{ccCrlc}[h!]
\tablecaption{\label{table.2}Predicted primordial light element abundances compared to the observational data for the case of a PMF with $\rho_{\rm Bc}/\rho_{\rm tot}=0.11$ and $\sigma_{\rm B}=0.53$.} 
\tablecolumns{8}
\tablenum{2}
\tablewidth{0pt}
\tablehead{
\colhead{Abundance} &
\colhead{$\eta_{10}=8.2$} &
\colhead{Observation}
}
\startdata
\bf{$Y_p$}                        & $0.2568$   & 0.2551$\pm$0.0022 \\
\bf{D/H($\times$10$^{5}$)}            & $3.21$  &3.02$\pm$0.23 \\
\bf{$^7$Li/H($\times$10$^{10}$)}     & $2.189$  & 1.70$\pm$0.64  \\
\enddata
\end{deluxetable*}
\\
\section{\label{sec:Conclusion}Conclusion}
In this work, an inhomogeneous PMF model is introduced during the BBN epoch, and has been explored. The PMF is described by a stochastic field constrained by the observed CMB power spectrum under the assumption of a power-law correlation function. However, the strength of the magnetic field varies spatially once the magnetic field is generated before weak decoupling. We adopt a PMF energy density characterized by a Gaussian dispersion in local field strength. This model implies the existence of an inhomogeneous PMF during the BBN epoch [Eq. (\ref{eq:16})]. We assume a homogeneous value of total energy density in the universe, and inhomogeneity of temperature along with that of the PMF. Locally, primordial baryons are in equilibrium with the same temperature which determines the photon energy density. Globally, due to the existence of an inhomogeneous PMF energy density, the temperature is inhomogeneous. This causes an effective non-MB distribution function for baryonic velocities during the BBN epoch. We derived an expression for the temperature distribution function [Eq. (\ref{eq:20})] and calculated the effective baryonic distribution function in our PMF model [Eqs. (\ref{eq:21})--(\ref{eq:19})]. We analyzed the reaction rates and concluded that charged particle reactions are affected most due to the Coulomb barrier while neutron induced reactions are not [Fig. \ref{fig7}]. The inhomogeneous PMF energy density was also added to the BBN network. We find that $^4$He abundance is most sensitive to $\rho_{\rm Bc}$ [Fig. \ref{fig3}]. We verified that under the limit of $\sigma_{\rm B} \to 0$, the abundances obtained from a homogeneous PMF strength are naturally recovered [Fig. \ref{fig3}].\\

In our model, the D and $^7$Li abundances are the most sensitive to the fluctuation parameter $\sigma_{\rm B}$ [Fig.\ref{fig4}]. By comparing our results with the $Y_p$ constraints, we find that $\rho_{\rm Bc}$ is less than $13\%$ of the total energy density, and the range of $\rho_{\rm Bc}/\rho_{\rm tot}=0.08-0.13$ provides the best fit to the observed abundances of for both $Y_p$ and D. This amount of magnetic energy density corresponds to a present PMF of $1.18-1.51\mu \rm G$. We conclude that the constraints from both $^4$He and D/H are satisfied with our PMF model for a fluctuation parameter $\sigma_{\rm B}=0.04-0.17$. Moreover, the $^7$Li abundance is reduced in our model to a value of $(3.35-3.52)\times10^{-10}$, which is still above the Spite plateau [Table \ref{table.1}]. If the baryon-to-photon ratio decreases from $\eta_{10} = 7.59-8.97$ during BBN to $\eta_{10} = 6.06-6.14$ of the Planck value by the time of photon last scattering, the Li problem could be solved for a fluctuation parameter of $\sigma_B=0.45-0.61$  [Table \ref{table.2}]. Such a high baryon-to-photon ratio does not result from the dissipation of PMF alone. However, if the present-day observed $^7$Li abundance level of the Spite plateau is the result of stellar depletion during the evolutionary stage of the metal-poor stars, this tension would be relaxed in our PMF model. There is another possibility of finding a change of the baryon-to-photon ratio by the radiative decay of exotic particles. Therefore, the above parameter region which we find in the inhomogeneous PMF model cannot be excluded at this time.
\\

\begin{acknowledgments}
YL was supported in part by Japan Foundation for Promotion of Astronomy. TK was supported by Grants-in-Aid for Scientific Research of JSPS KAKENHI (Grant Numbers JP15H03665 and JP17K05459). MK was partially supported by the visiting scholar program of NAOJ. Work of GJM supported in part by DOE nuclear theory grant DE-FG02-95-ER40934 and in part by the visitor program at NAOJ.
\end{acknowledgments}

\bibliographystyle{aasjournal}
\bibliography{PMF_reference}
\end{document}